%% file: main.tex
\documentclass{jpp}
\usepackage{graphicx}

\usepackage[utf8]{inputenc}
\usepackage[T1]{fontenc}
\usepackage{amsmath}

\usepackage{xspace}

\title{Reconstructing the Plasma Boundary with a Reduced Set of Diagnostics}

\author{
    M. S. Stokolesov\aff{1}\corresp{\email{ms@nextfusion.org}},
    M. R. Nurgaliev\aff{1},
    I. P. Kharitonov\aff{1},
    E. V. Adishchev\aff{1},
    D. I. Sorokin\aff{1},
    R. Clark\aff{2}
    \and D. M. Orlov\aff{2}
}

\affiliation{
\aff{1} Next Step Fusion, Luxembourg, Mondorf-les-Bains, 56A Avenue François Clément, L-5612
\aff{2} Center for Energy Research, University of California San Diego, La Jolla, CA 92122 USA
}

\begin{document}

\maketitle

\input{sections/new_commands}

\input{sections/abstract}

\input{sections/introduction}

\input{sections/dataset}

\input{sections/models}

\input{sections/results}

\input{sections/conclusion}

\input{sections/acknowledgements}

\bibliographystyle{jpp}
\bibliography{sections/biblist}

\end{document}

%% file: sections/new_commands.tex
\newcommand{\du}{\ensuremath{\delta_{\text{upper}}}\xspace}
\newcommand{\dl}{\ensuremath{\delta_{\text{lower}}}\xspace}
\newcommand{\Rg}{\ensuremath{R_{\text{geo}}}\xspace}
\newcommand{\Ru}{\ensuremath{R_{\text{upper}}}\xspace}
\newcommand{\Rl}{\ensuremath{R_{\text{lower}}}\xspace}
\newcommand{\Rmax}{\ensuremath{R_{\text{max}}}\xspace}
\newcommand{\Rmin}{\ensuremath{R_{\text{min}}}\xspace}

\newcommand{\Ncoils}{\ensuremath{N_\text{coils}}\xspace}
\newcommand{\Ip}{\ensuremath{I_\text{p}}\xspace}
\newcommand{\Vloop}{\ensuremath{V_\text{loop}}\xspace}

\newcommand{\Np}{\ensuremath{N_\text{p}}\xspace}
\newcommand{\Nc}{\ensuremath{N_\text{c}}\xspace}

\newcommand{\first}{\texttt{\#1}\xspace}
\newcommand{\second}{\texttt{\#2}\xspace}

%% file: sections/abstract.tex
\begin{abstract}
This study investigates the feasibility of reconstructing the last closed flux surface (LCFS) in the DIII-D tokamak using neural network models trained on reduced input feature sets, addressing an ill-posed task. Two models are compared: one trained solely on coil currents and another incorporating coil currents, plasma current, and loop voltage. The model trained exclusively on coil currents achieved a mean point displacement of $0.04$ m on a held-out test set, while the inclusion of plasma current and loop voltage reduced the error to $0.03$ m. This comparison highlights the trade-offs between input feature complexity and reconstruction accuracy, demonstrating the potential of machine learning algorithms to perform effectively in data-limited environments, such as those expected in Fusion Power Plants (FPP) due to diagnostic constraints imposed by the presence of blankets and shielding.
\end{abstract}

%% file: sections/introduction.tex
\section{Introduction}

In this study, we focus on reconstructing the plasma boundary, which is defined by the last closed flux surface (LCFS) of the confined plasma. The LCFS represents the transition between the closed magnetic field lines in the high-temperature, high-density confinement region of the plasma and the open field lines in the scrape-off layer of the plasma.

The position of the plasma boundary, as defined by the LCFS, is not directly measured during experiments. Instead, it is inferred through plasma equilibrium reconstruction using data collected from sensors installed in the tokamak. These calculations are performed by specialized equilibrium reconstruction codes, such as EFIT \citep{Lao_1985}, which is widely used for reconstructing equilibria in DIII-D discharges. In this work, we utilize EFIT data for training and evaluating machine learning (ML) models.

A significant limitation of traditional equilibrium reconstruction methods is their computational cost, particularly when multiple repeated calculations are required. Surrogate modeling provides a promising solution to address this issue by significantly reducing computation time while maintaining comparable accuracy.

There has been considerable progress in applying ML techniques to replace EFIT, most notably in the EFIT-AI project \citep{Lao_2022, Madireddy_2024}, which incorporates advanced methods and practices for reconstructing various plasma parameters, including the flux, toroidal current, and the full magnetic field distribution. However, the majority of existing works rely on comprehensive magnetics datasets, including signals from probes and flux loops, as input features for plasma boundary or full magnetic field distribution reconstruction \citep{Joung_2020, Wai_2022, Sun_2024}.

This reliance on a full set of magnetics data is common because it provides extensive constraints for plasma boundary reconstruction. Reducing the set of input parameters introduces additional challenges, as fewer constraints may lead to greater uncertainty in the reconstruction. Nevertheless, investigating models trained on limited diagnostic data is particularly relevant for data-limited environments, such as those anticipated in Fusion Power Plants (FPP), where diagnostic capabilities are constrained by the presence of blankets and shielding.

In this work, we explore the feasibility of reconstructing the LCFS in the DIII-D tokamak using neural network (NN) models trained on reduced feature sets. What differentiates our approach is the minimal input data used to train the models. Specifically, we compare two models: one trained exclusively on coil currents and another incorporating coil currents, plasma current, and loop voltage. This comparison illustrates the trade-offs between input feature complexity and reconstruction accuracy, highlighting the potential of machine learning algorithms to operate effectively in data-limited environments.

%% file: sections/dataset.tex
\section{Dataset}

The DIII-D discharge database encompasses a broad range of experimental data, including sensor measurements, EFIT equilibrium reconstructions, magnetic control system commands, and plasma state information. These data capture diverse experimental conditions, configurations, and control strategies. The data is recorded as high-resolution time series through an array of diagnostic systems installed in the tokamak, collecting key parameters such as magnetic fields, plasma temperature, and density. While coil currents $\{I_k\}_{k=1}^{\Ncoils}$, plasma current \Ip, and loop voltage \Vloop are measured during the discharge, the plasma geometry is automatically calculated post-shot using the EFIT equilibrium reconstruction code.

In this study, we consider the state of the plasma, which is defined by plasma shape, current, and kinetic profiles, and the state of the tokamak, which we define by coil currents. For brevity, we will occasionally use the term "discharge state" to refer to a state that describes both the plasma and the tokamak.

We employ neural network models based on Multi-Layer Perceptron (MLP) architectures, trained on historical EFIT data. The dataset consists of approximately 25,000 DIII-D discharges spanning 2004–2024, covering both positive (PT) and negative (NT) triangularities. The dataset was filtered to exclude discharges with missing EFIT signals, durations shorter than 1500 ms, or time steps exceeding 100 ms, resulting a final dataset of about 5 million discharge states used for training, validation, and testing.

Top and bottom triangularity (\du and \dl) characterize the deviation of the plasma boundary from an elliptical shape, where elongation ($\kappa$) defines the vertical stretching of the plasma and triangularity quantifies its indentation near the top and bottom. These parameters describe how much the plasma boundary deviates from an idealized elongated configuration and can be defined as:
\begin{equation*}
    \du = \frac{\Rg - \Ru}{a}\text{,}\quad
    \dl = \frac{\Rg - \Rl}{a}\text{,}
\end{equation*}
where $\Rg = \frac{\Rmax + \Rmin}{2}$ is the major radius of the plasma geometric center, \Ru and \Rl are the major radii of the highest and lowest points of the LCFS, $a = \frac{\Rmax - \Rmin}{2}$ is the plasma minor radius, \Rmax and \Rmin are the maximum and minimum values of the major radius along the LCFS.

The sign of \du and \dl distinguishes between positive and negative triangularity regimes. A \textit{positive triangularity} configuration ($\du > 0, \dl > 0$) corresponds to a plasma shape where the boundary is indented inward at the top and bottom, resulting in a D-shaped cross-section, which is commonly utilized in tokamaks to enhance plasma stability and confinement properties. Conversely, a \textit{negative triangularity} configuration ($\du < 0, \dl < 0$) features outward-extending top and bottom regions, yielding a \reflectbox{D}-shaped cross-section. Recent studies suggest that negative triangularity plasmas may exhibit reduced turbulence and improved confinement under certain operational conditions.

The DIII-D database contains many duplicated shots originating from repeated experiments. To ensure unbiased evaluation, it is critical to prevent duplicates from simultaneously appearing in the training, validation, or test sets. However, to our knowledge, no metadata exist that can reliably identify which shots belong to the same experiment or category, making duplicate identification challenging. To address this, we developed a data-driven method for detecting and removing duplicates. This approach involves two steps: first, each discharge is represented as a fixed-length sequence of discharge states; second, the discharges are clustered based on the similarity of these sequences. Using this method, we identified and removed approximately 30\% of the shots in the initial dataset as duplicates.

The signals in the DIII-D database are recorded as time series, typically with different timestamp resolutions. While these signals span the same time intervals, they are often sampled at varying discretization rates. To align the signals along a common timestamp dimension, we interpolated all signals (i.e., coil currents, \Ip, and \Vloop) to match the timestamps of the plasma boundary signal for each discharge.

As described earlier, the plasma boundary represents the outermost surface of the confined plasma and is typically defined as a two-dimensional curve. The DIII-D database provides a discretized version of this curve as a set of $(R, Z)$ points in the tokamak's coordinate system for each timestamp in a discharge. Since the plasma boundaries from different discharges are described by varying numbers of points, we interpolated all boundaries to a fixed number of points using a polar representation of the plasma shape. Specifically, we converted the $(R, Z)$ points to polar coordinates relative to the magnetic center of the plasma. The boundary was then uniformly sampled across $90$ polar angles to produce a fixed-dimensional representation in the $(R, Z)$ coordinate system for each timestamp.

%% file: sections/models.tex
\section{Models}

To investigate the feasibility of reconstructing the plasma boundary using a reduced set of input features, we compared the performance of two ML models trained on the same set of discharge states but with different input features (Table \ref{tab:models}). Each model’s task is to reconstruct the plasma boundary using the corresponding set of input features at the same timestamp. The models do not rely on information from prior states of the plasma or device and thus perform the same function as the EFIT code.

\begin{table}
  \begin{center}
  \begin{tabular}{c@{\hspace{0.5cm}}c@{\hspace{0.5cm}}c@{\hspace{0.5cm}}c}
      Model & $\{I_k\}_{k=1}^{\Ncoils}$ & \Ip & \Vloop \\[3pt]
      \texttt{\#1} & \checkmark & - & - \\
      \texttt{\#2} & \checkmark & \checkmark & \checkmark \\
  \end{tabular}
  \caption{Correspondence between models and the input feature sets they were trained on. Here, $k$ represents the magnetic coil index, and $\Ncoils = 20$ in this work for DIII-D.}
  \label{tab:models}
  \end{center}
\end{table}

Model \first was trained to reconstruct the plasma boundary using only the coil current values as input. Model \second extended the input feature set to include two additional parameters: the plasma current \Ip and the loop voltage \Vloop.

The DIII-D tokamak has 18 shaping coils (F-coils) and 6 ohmic heating coils (E-coils). In this work, we use only two of the E-coils, "ECOILA" and "ECOILB", which together form the center solenoid. The remaining four E-coils share the same power supply as these two and thus exhibit highly correlated values. As a result, we have a total of 20 coil current values, denoted as $\{I_k\}_{k=1}^{20}$, which form the input vector for model \first. The input for model \second consists of 22 values, $\{I_k\}_{k=1}^{20} \cup \{\Ip, \Vloop\}$.

Both models predict the geometry of the plasma boundary as a vector of size $\Nc = 2 \times \Np$. This vector represents a flattened matrix of shape $(\Np, 2)$, where $\Np$ corresponds to the number of 2D points describing the plasma boundary at a given timestamp, which is set to $90$ in our experiments.

The models are designed to receive the state of a discharge at a specific moment in time and compute the corresponding plasma boundary for that moment. Consequently, the training dataset comprises individual discharge states and is represented as a matrix of shape $(N, D + \Nc)$, where $N$ is the total number of discharge states, $D = 20 \text{ or } 22$ corresponds to the dimensionality of the input feature space, and $\Nc = 180$ represents the dimensionality of the plasma boundary vectors. Before being used for training, both the input features and the target outputs are standardized separately by subtracting their respective means and dividing by their standard deviations.

\begin{figure}
  \centering
  \includegraphics[scale=0.145]{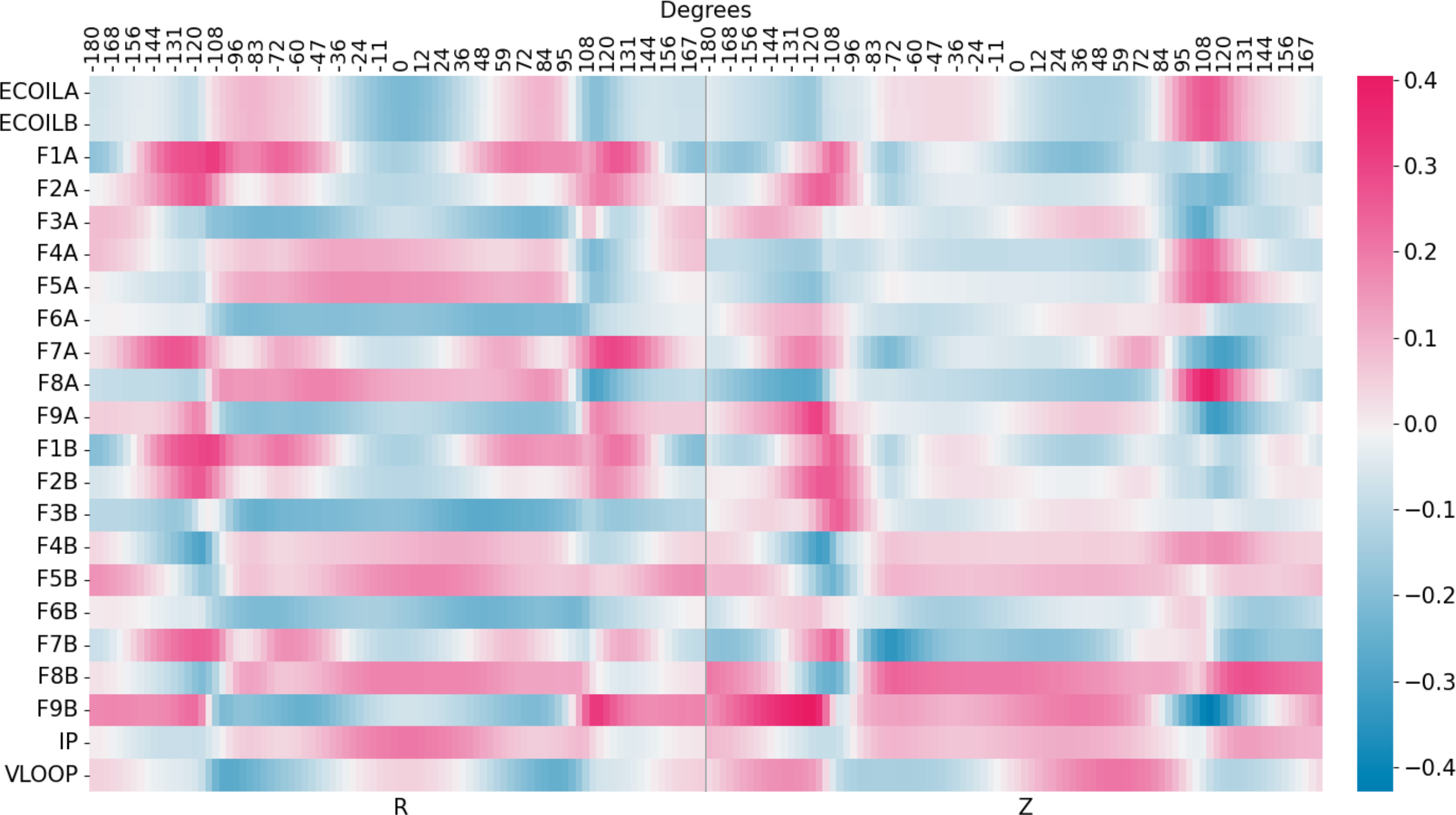}
  \caption{Heatmap of the correlation matrix between dataset features (model inputs, left) and target plasma boundary points (model outputs, top). The left side of the matrix corresponds to the $R$ coordinates of the boundary points, the right side corresponds to the $Z$ coordinates. For clarity, each boundary point is labeled by its polar angle (in degrees) relative to the plasma center and the positive direction of the $R$ axis.}
\label{fig:corr_heatmap}
\end{figure}

Figure \ref{fig:corr_heatmap} shows a correlation matrix between the dataset features (model inputs) and the target plasma boundary points (model outputs). The matrix highlights small but noticeable correlations, ranging from approximately $0.2$ to $0.4$, between certain coil currents and specific boundary points (both $R$ and $Z$ coordinates), providing insights into how variations in input features correspond to changes in the plasma boundary.

Both models share the same fully connected neural network (FCNN) architecture, consisting of two hidden layers with 150 and 80 neurons, respectively. To stabilize training and prevent overfitting, we apply batch normalization \citep{Ioffe_2015} and dropout (probability $0.2$) \citep{Srivastava_2014}. ReLU is used as the activation function.

To train the models, we employed the mean squared error (MSE) loss function, which has demonstrated strong performance in predicting plasma parameters \citep{Abbate_2021, Wan_2024}:
\begin{equation*}
    \text{MSE\,Loss} = \frac{1}{N} \sum_{i=1}^N \left\|b_i - \hat{b}_i\right\|^2\text{,}
\end{equation*}
where $b_i$ is the ground truth boundary vector, $\hat{b}_i$ is the predicted boundary vector, $i$ is the sample index, and $N$ is the size of the training dataset. We used a learning rate of $1 \cdot 10^{-4}$ and the Adam \citep{Kingma_2014} optimizer for training.

To prevent data leaks, we adopted a shape-based cross-validation approach. Discharge states were first divided into three main groups based on top and bottom triangularity; each group was then further subdivided into four subgroups according to median values. This process produced 12 distinct sets of plasma shapes. At each cross-validation stage, the training–validation set consisted of 11 of these groups, while the remaining group was used for testing (figure \ref{fig:cv_results}). The validation set at each step was drawn from states belonging to discharges in 2020, 2021 and 2022. To prevent overfitting, we balanced the number of states in each group by subsampling, ensuring that every plasma shape group contained the same number of states. As a result, the training–validation–test ratio was approximately 80–10–10 at each cross-validation stage.

%% file: sections/results.tex
\section{Results}

\begin{figure}
  \centering
  \includegraphics[scale=0.31]{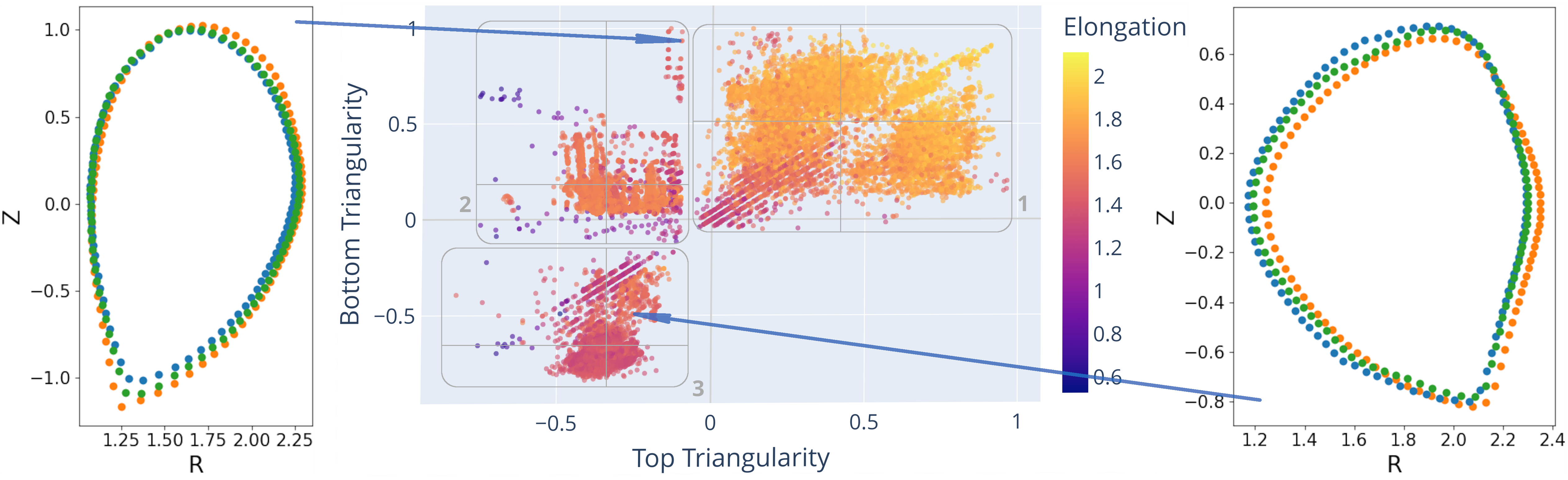}
  \caption{Center plot: Distribution of plasma states in the training dataset based on top and bottom triangularity. All states are divided into three main groups, producing 12 subgroups for cross-validation. At each cross-validation stage, the training set is composed of data from 11 groups, while the remaining group is used for testing. The left and right plots show examples of plasma shapes with "negative-positive" and "negative-negative" triangularity, respectively, illustrating the models' accuracy on previously unseen cases (orange -- true boundary, blue -- model \first reconstruction, green -- model \second).}
\label{fig:cv_results}
\end{figure}

To evaluate model performance during cross-validation and on the test set, three metrics were used:
\begin{itemize}
    \item[--] \textit{Maximum Point Displacement} (MXD) and \textit{Mean Point Displacement} (MND):
    \begin{equation*}
        \text{MXD} = \frac{1}{N} \sum_{i=1}^{N} \max_{j=1}^{\Np}\left\|p_{ij} - \hat{p}_{ij}\right\|\text{,}
    \end{equation*}
    \begin{equation*}
        \text{MND} = \frac{1}{N} \sum_{i=1}^{N} \frac{1}{\Np} \sum_{j=1}^{\Np} \left\|p_{ij} - \hat{p}_{ij}\right\|\text{,}
    \end{equation*}
    where $p_{ij}$ and $\hat{p}_{ij}$ are the $j$-th true and predicted 2D boundary points for the $i$-th sample, \Np is the number of boundary points, and $N$ is the total number of samples. Hereafter, all MXD and MND values are reported in meters and calculated in the original scale of the quantities.
    
    \item[--] \textit{Coefficient of Determination} $R^2$:
    \begin{equation*}
        R^2 = \frac{\sum_{i=1}^{\Nc}\left(V_i \cdot R_i^2\right)}{\sum_{i=1}^{\Nc} V_i}\text{,}\quad
        V_i = \frac{1}{N} \sum_{j=1}^{N} \left(y_{ij} - \overline{y_{i}}\right)^2\text{,}\quad
        R_i^2 = 1 - \frac{\sum_{j=1}^{N}\left(y_{ij} - \hat{y}_{ij}\right)^2}{\sum_{j=1}^{N}\left(y_{ij} - \overline{y_{i}}\right)^2}\text{,}
    \end{equation*}
    where $R_i^2$ is the coefficient of determination for the $i$-th network output (corresponding to the $R$ or $Z$ coordinate of a boundary point), $\Nc = 180$ (as $90$ points are represented by both $R$ and $Z$ coordinates), $y_{ij}$ and $\hat{y}_{ij}$ are the true and predicted values of the $i$-th coordinate for the $j$-th sample, $\overline{y_i}$ is the mean true value of the $i$-th coordinate, $N$ is the number of samples, and $V_i$ reflects variance in $y_i$.
\end{itemize}

\begin{table}
  \begin{center}
  \begin{tabular}{c@{\hspace{0.5cm}}c@{\hspace{0.5cm}}c@{\hspace{0.5cm}}c}
      Model & $R^2$ & MXD & MND \\[3pt]
      \texttt{\#1} & $0.52$ & $0.10$ & $0.03$ \\
      \texttt{\#2} & $0.57$ & $0.08$ & $0.02$ \\
  \end{tabular}
  \caption{
  Performance of the models during cross-validation. The metric values are averaged across cross-validation splits.
  }
  \label{tab:cv_results}
  \end{center}
\end{table}

The results of cross-validation are presented in table \ref{tab:cv_results}, which summarizes the performance of the two models across all splits. For each split, the testing set contained unseen plasma shapes, ensuring that the models were evaluated on data outside their training subsets. In our experiments, the trained models \first and \second achieved cross-validated $R^2$ scores of $0.52$ and $0.57$ respectively. Although these coefficients of determination indicate that the models capture some variability in the shape, they also tell that a significant amount of the underlying relationship remains unexplained, in part due to the reduced set of features used. The metrics show that model \second, which incorporates coil currents, \Ip and \Vloop, achieves higher accuracy compared to the coil-current-only model, highlighting the added value of including additional input features. 

\begin{figure}
  \centering
  \includegraphics[scale=0.58]{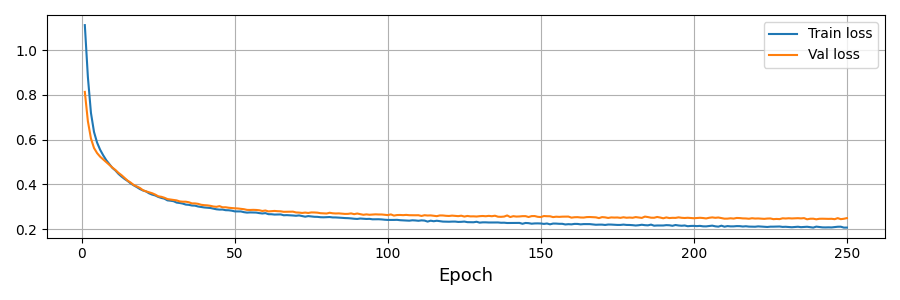}
  \caption{
    Dynamics of the MSE loss during training of the model \second for the training set (blue) and the validation set (orange).
  }
\label{fig:losses}
\end{figure}

The test subset included discharges conducted in 2024. Similar to the cross-validation procedure, the validation set for the final models was drawn from states belonging to discharges in 2020, 2021 and 2022. To prevent overfitting and ensure accurate model evaluation, we balanced the number of states corresponding to different plasma shapes in the training and testing sets using a subsampling approach similar to that used for cross-validation. As a result, the training–validation–test ratio was approximately 75–10–15. The dynamics of the loss functions during the training of the model \second are shown in figure \ref{fig:losses}.

\begin{table}
  \begin{center}
  \begin{tabular}{c@{\hspace{0.5cm}}c@{\hspace{0.5cm}}c@{\hspace{0.5cm}}c}
      Model & $R^2$ & MXD & MND \\[3pt]
      \texttt{\#1} & $0.55$ & $0.12$ & $0.04$ \\
      \texttt{\#2} & $0.63$ & $0.11$ & $0.03$ \\
  \end{tabular}
  \caption{
  Performance of the models on the test set.
  }
  \label{tab:test_results}
  \end{center}
\end{table}

The performance of both models on the test subset is summarized in table \ref{tab:test_results}. The results show that the models differ in performance by approximately $0.01$ m in favor of model \second based on the MXD and MND metrics. Model \second, which used coil currents, \Ip, and \Vloop as input, achieved an MND of $0.03$ m, demonstrating the feasibility of reconstructing the LCFS with a reduced diagnostic set. Standard DIII-D equilibrium reconstruction techniques and the plasma control system ensure accuracies of about $10^{-2}$ m required for experimental programs \cite{Eldon_2020}. When using a model trained on the reduced set of diagnostics, the error increases by only $0.01$-$0.02$ m, which is a promising outcome for FPP applications.

\begin{figure}
  \centering
  \includegraphics[scale=0.58]{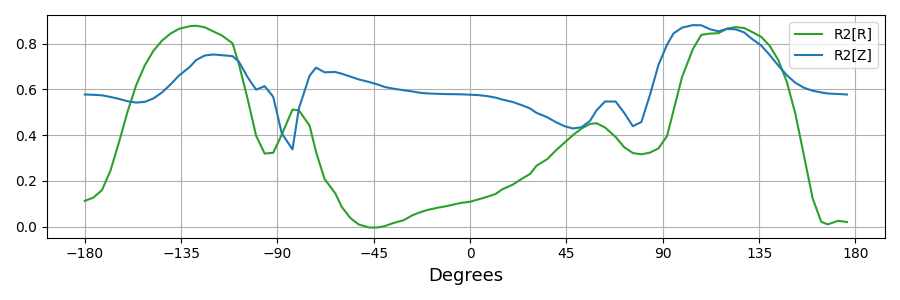}
  \caption{
    Distribution of the $R^2$ metric for the $R$ and $Z$ coordinates of plasma boundary points for the model \second on the test set. Plasma boundary points are labeled by their polar angle (in degrees) relative to the plasma center and the positive direction of the $R$ axis.
  }
\label{fig:r2_dist}
\end{figure}

A more detailed assessment of the model \second performance on the test subset can be obtained from the distribution of the $R^2$ metric across all plasma boundary points. As shown in figure \ref{fig:r2_dist}, $R^2$ values are distributed unevenly across different regions of the plasma boundary. The most likely explanation for this behavior is that, due to the use of a reduced set of input features, some regions of the plasma boundary exhibit very weak dependence on the available inputs. As a result, the MLP model struggles to achieve significantly better performance than the mean model in these areas. This uneven relationship between the input features and different regions of the plasma boundary is also evident from the correlation matrix shown in figure \ref{fig:corr_heatmap}.

\begin{figure}
  \centering
  \includegraphics[scale=0.085]{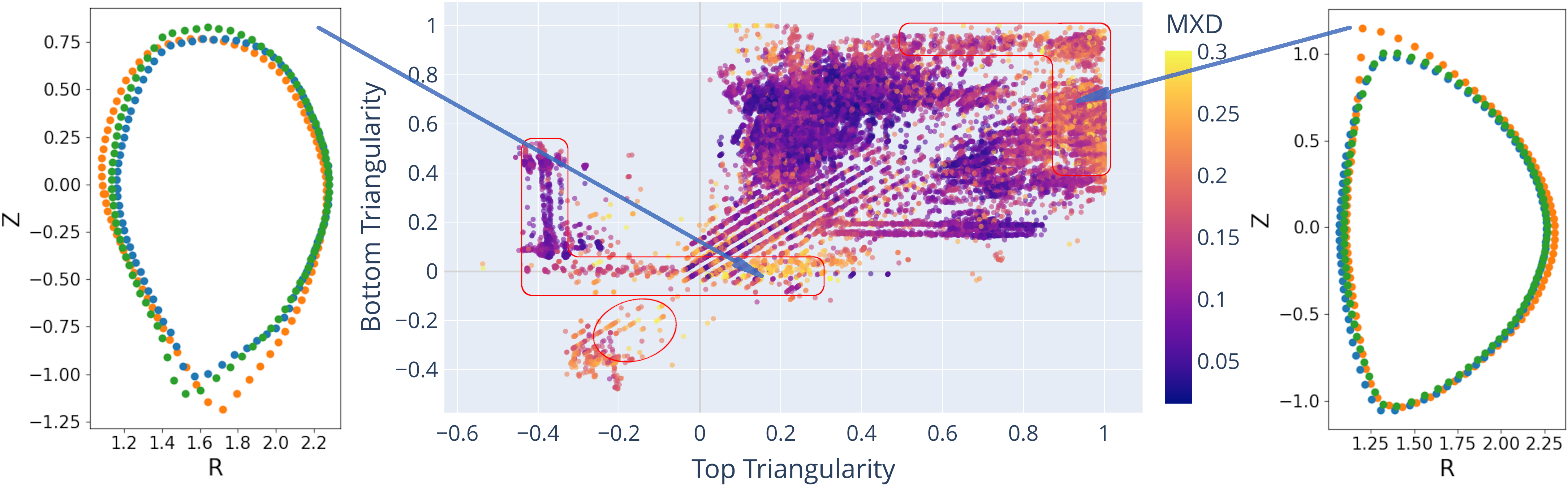}
  \caption{
    Center plot: Distribution of the MXD metric across plasma states in the test subset (model \second). Examples of plasma shapes are shown for "positive-negative" triangularity (left plot) and "positive-positive" triangularity (right plot), with orange indicating the true boundary, blue -- model \first reconstruction, and green -- model \second. These examples are taken from regions (highlighted in red) far from the majority of training samples (see figure \ref{fig:cv_results}) and represent challenging cases for the models (MXD values of $0.25$-$0.3$ m).
  }
\label{fig:test_results}
\end{figure}

\begin{figure}
  \centering
  \includegraphics[scale=1.2]{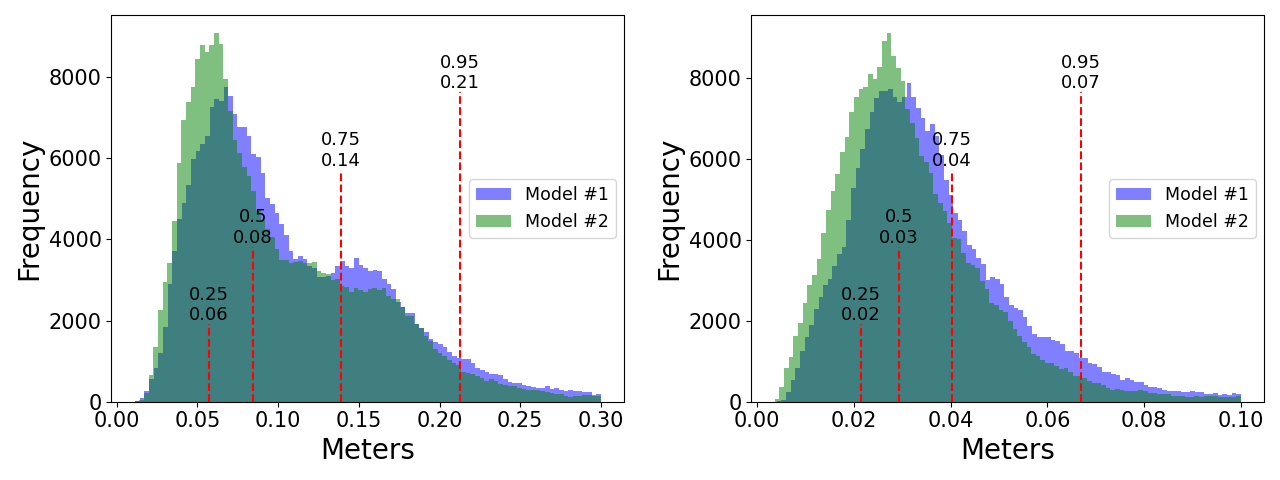}
  \caption{
  Comparison of MXD (left) and MND (right) metric distributions for models \first (blue) and \second (green) on samples from the entire test subset. Histogram values are computed for individual samples. Vertical red lines indicate quantiles for model \second, with the top number representing the quantile and the bottom number showing the corresponding quantile value.
  }
\label{fig:mxd_mnd_hists}
\end{figure}

The distribution of model errors across the entire test subset is shown in figure \ref{fig:test_results}. To further evaluate model performance on the test subset, we plotted histograms of the MXD and MND metrics across all test samples in figure \ref{fig:mxd_mnd_hists}. These histograms show that the error distribution for model \second is shifted towards lower values compared to the model \first.

It is important to note that, in general, there cannot be a one-to-one correspondence between coil currents and plasma shape. The coil-current-only model appears to have learned to reconstruct plasma shapes based on currents generated by specific control algorithms. This observation underscores the critical role of representative training data in machine learning approaches for plasma boundary reconstruction.

%% file: sections/conclusion.tex
\section{Conclusion}

This study investigates the performance of reconstructing the plasma boundary using neural network models trained on reduced input feature sets. Specifically, we compare two models: one relying solely on coil currents, and another that also incorporates plasma current and loop voltage. The coil-current-only model achieves a mean point displacement of 0.04 m on a held-out test set, while adding plasma current and loop voltage reduces the error to 0.03 m, demonstrating the impact of these features on reconstruction quality. The results also show, that even with minimal diagnostics, it is feasible to reconstruct the plasma boundary, although the models capture only part of the plasma shape's variability. These results are promising for future applications in Fusion Power Plants (FPP), where diagnostic capabilities will be constrained by the presence of blankets and shielding.

Future work will focus on evaluating the generalizability of these models to out-of-distribution data and further exploring their applicability to FPP environments. This includes extending the feature set to incorporate additional diagnostics and testing the models on synthetic datasets that emulate the conditions in next-generation fusion devices.

%% file: sections/acknowledgements.tex
\section{Acknowledgements}
This material is based upon work supported by the U.S. Department of Energy, Office of Science, Office of Fusion Energy Sciences, using the DIII-D National Fusion Facility, a DOE Office of Science user facility, under Award(s) DE-FC02-04ER54698 and by Next Step Fusion funding. The authors would like to thank Dr. Cihan Akçay and Dr. Scott E. Kruger for helpful discussions and valuable insights, which greatly contributed to this work.

This report was prepared as an account of work sponsored by an agency of the United States Government. Neither the United States Government nor any agency thereof, nor any of their employees, makes any warranty, express or implied, or assumes any legal liability or responsibility for the accuracy, completeness, or usefulness of any information, apparatus, product, or process disclosed, or represents that its use would not infringe privately owned rights. Reference herein to any specific commercial product, process, or service by trade name, trademark, manufacturer, or otherwise does not necessarily constitute or imply its endorsement, recommendation, or favoring by the United States Government or any agency thereof. The views and opinions of authors expressed herein do not necessarily state or reflect those of the United States Government or any agency thereof.